\documentclass[3p]{elsarticle}
\usepackage{amssymb}
\usepackage{amsgen}
\usepackage{amsfonts}
\usepackage{amsbsy}
\usepackage{color}
\usepackage{natbib}
\usepackage{tikz} 
\usepackage{graphicx}				
\usepackage{amsmath}
\usepackage{algorithm}
\usepackage{algpseudocode}

\usetikzlibrary{calc}				
\usepackage{amssymb}
\usepackage{pgfplots}
\usetikzlibrary{calc,angles,positioning,intersections,quotes,decorations.markings}
\usepackage{graphicx}

\usepackage{amsfonts}
\usepackage{times}
\usepackage{latexsym}
\usepackage{amssymb}
\usepackage{amsmath}
\usepackage{verbatim}
\newtheorem{theorem}{Theorem}
\newtheorem{assumption}[theorem]{Assumption}
\newtheorem{corollary}[theorem]{Corollary}
\newtheorem{rem}{Remark}
\newtheorem{definition}[theorem]{Definition}
\newtheorem{example}[theorem]{Example}
\newtheorem{lemma}[theorem]{Lemma}
\newenvironment{proof}{ \textbf{Proof:} }{ \hfill $\Box$}
\newtheorem{prop}{Proposition}

\def\bb0{{\mathbb{0}}}

\def\bb{{\mathbf{b}}}

\def\bg{{\mathbf{g}}}

\def\bp{{\mathbf{p}}}
\def\bq{{\mathbf{q}}}

\def\b0{{\mathbf{0}}}
\def\opt{\mathsf{OPT}}

\def\b1{{\mathbf{1}}}

\def\bbE{{\mathbb{E}}}

\def\bbP{{\mathbb{P}}}

\def\bbR{{\mathbb{R}}}

\def\cA{\mathcal{A}}
\def\cB{\mathcal{B}}

\def\cD{\mathcal{D}}
\def\cE{\mathcal{E}}

\def\cP{\mathcal{P}}

\def\cR{\mathcal{R}}

\def\cT{\mathcal{T}}

\def\cW{\mathcal{W}}

\def\sfa{{\mathsf{a}}}
\def\sfb{{\mathsf{b}}}
\def\sfc{{\mathsf{c}}}

\def\sfm{{\mathsf{m}}}

\def\sfv{{\mathsf{v}}}

\def\sf0{{\mathsf{0}}}

\def\nn{\nonumber}

\begin{document}
	\begin{frontmatter}
		\title{Online Bidding Algorithms with Strict Return on Spend (ROS) Constraint}
		\author[rahul]{Rahul Vaze}
		\ead{rahul.vaze@gmail.com}
		\author[abhishek]{Abhishek Sinha}
		\ead{abhishek.sinha@tifr.res.in}
		\address[rahul]{School of Technology and Computer Science, Tata Institute of Fundamental Research, Mumbai, India}
		\address[abhishek]{School of Technology and Computer Science, Tata Institute of Fundamental Research, Mumbai, India}		
		\journal{Performance Evaluation (PEVA)}
		\date{}
		\begin{abstract}%
			Auto-bidding problem  under a strict return-on-spend constraint (ROSC) is considered, where an algorithm has to make decisions about how much to bid for an ad slot depending on the revealed value, and 
the hidden allocation and payment function that describes the probability of winning the ad-slot depending on its bid. The objective of an algorithm is to maximize the expected utility (product of ad value and probability of winning the ad slot) summed across all time slots subject to the total expected payment being less than the total expected utility, called the ROSC. 
A (surprising) impossibility result is derived that shows that no online algorithm can achieve a sub-linear regret even when the value, allocation and payment function are drawn i.i.d. from an unknown distribution.
The problem is non-trivial even when the revealed value remains constant across time slots, and an algorithm with regret guarantee that is optimal up to logarithmic factor is derived.

		\end{abstract}
	\end{frontmatter}


\begin{abstract} 

\end{abstract}


\section{Introduction}
In this paper, we consider the {\it auto-bidding} problem, an important example of online optimization under time varying constraints, that is relevant for many resource allocation settings, especially for online advertising. 
With auto-bidding, in each time slot $t$, an auction is conducted, where 
an ad-query defined by a three-tuple, {\it value} $v_t$, {\it allocation function} $x_t(.)$ and {\it payment function} $p_t(.)$ is generated that is independent and identically distributed (i.i.d.) with some unknown distribution. 
Allocation function $x_t(b_t)$ is essentially the probability of winning the auction in slot $t$ depending on the bid value $b_t$. 
Corresponding to the allocation function is the expected payment $p_t$ that the bidder has to pay if allocation function is non-zero for the submitted bid. Both the allocation and payment functions are assumed to be monotonic. 

In each slot, a single auto-bidder gets to see only the value  and is asked to bid, while the allocation and payment functions remains {\bf hidden}. 
Once the bid is submitted, the bidder receives the outcome of the auction: the hidden allocation function and the required payment. The realized value for the bidder in a slot is the product of the {\it value} and the {\it allocation function} evaluated at the submitted bid for that slot. 
Because of financial considerations, a typical return-on-spend  constraint (ROSC) is imposed that requires that the {\it total expected realized value summed across all slots} should be at least as much as the {\it total expected payment made across all slots}. 

The optimization problem is to maximize the total expected realized value summed across all slots subject to the ROSC. The problem is {\it online}, since at any time slot an online algorithm can only use the revealed value of that slot, and all the past information. Regret
$\cR_\cA(T)$ quantifies the performance of any online algorithm $\cA$, defined as the expected difference between the total realized value of an optimal offline algorithm $\opt$ and  $\cA$, over a time horizon of $T$ slots.

Auto-bidding is the dominant model in online advertising \cite{balseiro2019learning, aggarwal2019autobidding, babaioff2020non, golrezaei2021auction, deng2021towards, balseiro2021robust, balseiro2021landscape}, where auto-bidders are deployed on behalf of advertisers to win ad-slots on webpages.
Once a user arrives on a webpage, the effective value of showing an ad to that user is some fixed basic value (intrinsic to the user) multiplied by the probability of an advertiser winning that auction depending on its bid, and the associated payment rule that is required to be truthful for maintaining 
incentive compatibility.

\subsection{Prior Work}
Single learner auto-bidding problem has been considered quite extensively in literature, but one allowance that is almost universally made is accepting some violation of the ROSC. For example, in \cite{Feng}, an algorithm with $\cR(T) = O(\sqrt{T})$ is proposed but it violates the ROSC by an amount  $O(\sqrt{T\log T})$. Prior to this, \cite{castiglioni2022unifying} derived an algorithm with $\cR(T) =O(T^{3/4})$ together with $O(T^{3/4})$ ROSC violation. For a slightly different objective function of $(v_t-s_t)$ at time $t$, where $s_t$ is the second largest bid at time $t$ from other bidders, and both $v_t,s_t$ are generated i.i.d., \cite{golrezaei2021bidding} derives a threshold based algorithm with $\cR(T) = O(T^{2/3})$ and satisfies the ROSC. In addition to ROSC, additional constraint on total payment made 
has also been considered in \cite{Feng, castiglioni2022unifying, golrezaei2021bidding}. 
Related to auto-bidding problem is the online allocation problem \cite{balseiro2020dual}, that only has budget constraints and no ROSC.

More complicated auto-bidding models \cite{borgs2007dynamics, golrezaei2021auction, chen2023complexity, lucier2024autobidders} 
have also been studied where multiple agents compete directly, however, always following a second-price auction, for which $O(\sqrt{T})$ regret and ROSC satisfaction with high probability are known.

\subsection{Related Problems} 
In constrained online convex optimization (COCO), on each round, the online policy first chooses an action, and then the adversary reveals a convex cost function and multiple convex constraints. Since constraints are revealed after the action has been chosen, no policy can satisfy the constraints exactly, and hence
in addition to regret, cumulative constraint violation (CCV) is a relevant performance metric. COCO has been studied with both 
 stochastic \cite{mannor2009online, mahdavi2013stochastic, yu2017online, badanidiyuru2018bandits} as well as worst case input \cite{yu2016low, guo2022online, sinha2024optimalalgorithmsonlineconvex}, where simultaneous regret and CCV bounds are derived.
Compared to COCO, with auto-bidding,  value $v_t$ is revealed before action is taken and the input is i.i.d.
 
 Other related models of online optimization with constraints have also been studied, e.g. \cite{agrawal2014fast}, where at each slot a discrete set of vectors are revealed, and an algorithm has to choose one of them with the goal of minimizing a (pre-specified) loss function evaluated at the average of the chosen vectors, subject to the average of the chosen vectors to lie in a (pre-specified) convex region. Similar to COCO, simultaneous regret and constraint violation bounds are derived in \cite{agrawal2014fast}.
 
 {\bf Bandits with Knapsack (BwK)} Bandits with knapsacks is a popular model \cite{badanidiyuru2018bandits, rangi2018unifying, immorlica2022adversarial, rivera2025online}, where there are $K$ arms, and at each time slot only one arm can be pulled/sampled by an algorithm. Pulling an arm reveals the associated reward, and a $d$-dimensional non-negative resource consumption vector. For each resource $i=1, \dots, d$, there is a  budget constraint $B_i$, and the algorithm can sample any arm for only as long as the cumulative consumption for any of the $i=1, \dots, d$ resources is at most its constraint $B_i$. The objective is to find an algorithm that maximizes the cumulative reward. For BwK, when the reward and consumption vector are stochastic with unknown distribution, algorithms with regret $O(\sqrt{T})$  are known \cite{badanidiyuru2018bandits, rangi2018unifying, immorlica2022adversarial, rivera2025online}, when $B_i =\Omega(\sqrt{T})$ for all $i$, and $T$ is the time horizon. The adversarial setting is also considered for BwK where competitive ratio guarantees of $\log(T)$ have been obtained \cite{immorlica2022adversarial, rivera2025online}.
 
 The connection of stochastic BwK with the considered auto-bidding can be understood as follows.  For the auto-bidding problem, at time $t$, once the value $v_t$ is revealed, consider an infinite 
 number of arms corresponding to possible bids than an algorithm can make. For an arm that corresponds to bid $b$, the associated reward of that arm is $v_t\cdot x_t(b)$ and the consumption is $x_t(b)(v_t - p_t(b))$, the expected utility minus the expected payment. Moreover, the ROSC corresponds to overall consumption constraint of $B=0$, but the consumption $x_t(b)(v_t - p_t(b))$ at time $t$ is possibly negative as well. 
The fact that the consumption can be both positive and negative unlike BwK, an algorithm for auto-bidding does not have to necessarily terminate at any time (budget exhaustion time for BwK) earlier than the end of time horizon.
 Thus, there are basic conceptual 
 differences between Stochastic BwK and the considered auto-bidding problem, which result in fundamentally different technical results as will be described next.

{\bf Knapsack Problem:} The auto-bidding problem is closely related to an interesting variant (unstudied as far as we know) of the usual online knapsack problem \cite{vaze2023online}, where on an item's arrival, only its value is revealed 
but the size (it occupies) is hidden. An algorithm is allowed to advertise the maximum size it is willing to accept knowing the value, and if the size is at most the advertised size, the item is accepted.
If the input is worst-case, it is not difficult to see that for this variant of the knapsack problem, the competitive ratio of any online algorithm is unbounded. Thus, the auto-bidding problem's special structure and the i.i.d. input makes it amenable for non-trivial analysis. Knapsack problem with unknown but fixed capacities has been studied \cite{disser2017packing}, but not under  above constraints. 

{\bf Resource Allocation:} The single auto-bidder problem is essentially a resource allocation problem,
where a decision maker needs to
choose an action that consumes a certain amount
of resources and generates utility/revenue. The revenue
function and resource consumption of each request are drawn i.i.d. from an unknown distribution. 
The objective is to maximize
cumulative revenue subject to some constraint on the total consumption of resources. 
Thus, the considered problem is relevant for very general settings that include 
inventory management for hotels and airlines with limited resources \cite{talluri1998analysis}, buying energy/power under 
i.i.d. demand and pricing \cite{yang2019online}, etc.

\subsection{Our Contributions}
In this paper, we consider the single auto-bidding problem with i.i.d. input, where in contrast to prior work that allowed ROSC 
violation, the main object of interest  is the characterization of achievable regret under strict ROSC satisfaction. Towards that end, our contributions are as follows.
\begin{itemize}
\item We show a {\bf major negative result} that the regret of any online algorithm that strictly satisfies the ROSC is $\Omega(T)$. Thus imposing strict ROSC satisfaction makes 
all online algorithms `weak' for auto-bidding, and to {\bf get sub-linear regret, ROSC has to be violated}. What is the best simultaneous regret and ROSC violation that any online algorithm can achieve remains an open question. $O(\sqrt{T})$ regret and $O(\sqrt{T \log T})$ are currently the best known simultaneously achievable regret and ROSC violation bound \cite{Feng}.
\item For the case of repeated identical auction when value ($v_t$) revealed in each slot is a constant, we show that the sum of the regret and the ROSC 
violation for any online algorithm is $\Omega(\sqrt{T})$. This implies that  the regret of any online algorithm that strictly satisfies the ROSC is $\Omega(\sqrt{T})$.
\item We show that the best known algorithm \cite{Feng} for the considered auto-bidding problem which violates the ROSC by at most $O(\sqrt{T\log T})$ has regret $\Omega(\sqrt{T})$ even for repeated identical auction. Thus the $O(\sqrt{T})$ regret bound of \cite{Feng} is tight.

\item For the repeated identical auction when value ($v_t$) revealed in each slot is a constant, we propose a simple algorithm that satisfies the ROSC on a sample path basis and show that its regret is at most $O(\sqrt{T \log T})$ for a specific class of allocation and payment functions under some mild assumptions. As far as we know, this is first such result.

\item Since achieving sub-linear regret is not possible for auto-bidding when $v_t\ne v_{t'}$,  
we consider the metric of competitive ratio and propose an algorithm with competitive ratio close to $1/2$ for a specific class of allocation and payment functions.
\end{itemize}

\section{Problem Formulation}
We consider the following online learning model, where time is discrete and in time slot $t$: an ad query is generated that has three attributes, its value $v_t \in [0,1]$,  the allocation function $x_t: \bbR^+ \cup \{0\} \rightarrow [0,1]$, and the payment function $q_t: \bbR^+ \cup \{0\}\rightarrow [0,1]$. 
The allocation function $x_t(b_t)$ is effectively the probability of `winning' the slot $t$ by bidding $b_t$, and the expected valuation accrued from slot $t$ is $v_t x_t(b_t)$. The allocation function $x_t$ is required to be a non-decreasing function of the bid $b_t$ with $x_t(0) =0$. The payment function $q_t$   is also non-decreasing, is zero when the allocation is zero, and is at most the bid. Thus, the expected payment function is $p_t(b_t)= q_t(b_t)x_t(b_t)$ which for $b_t=v_t$ satisfies
 \begin{equation}\label{eq:paymentcondition}
p_t(v_t) \le v_t x_t(v_t).
\end{equation}

Instead of working with $q_t$, we work with $p_t$ from here onwards.
The value $v_t$ is visible to the online learner before making a decision at time slot $t$ about the bid $b_t$ it wants to make. Once $b_t$ is chosen, both $x_t(b_t)$ and $p_t(b_t)$ are revealed. 
We consider the stochastic model, where the input $(v_t,x_t,p_t)$ is generated i.i.d. from an unknown distribution $\cD$. 

The problem for the online learner is to choose $b_t$ in slot $t$ knowing the past information $$\{(v_\tau, x_\tau(b_\tau), p_\tau(b_\tau))_{\tau=1, \dots, t-1}\}$$ and the present value $ v_t$ so as to maximize the total expected accrued valuation subject to ROSC, {\it that requires that the 
total expected payment should be at most the total accrued valuation}. Formally, the online problem is 
\begin{align}\nn
\max_{b_t} & \quad V(T)=  \bbE\left\{\ \sum_{t=1}^T v_t x_t(b_t)\right\},\\\label{eq:prob}
\text{subj. to} &  \quad \bbE\left\{\sum_{t=1}^Tp_t(b_t)\right\} \le   \bbE\left\{\ \sum_{t=1}^T v_t x_t(b_t)\right\}.
\end{align}
where constraint in \eqref{eq:prob} is the ROSC. 
We refer to our {\bf benchmark} as the $\opt$, that solves \eqref{eq:prob} knowing $\cD$, the distribution for $(v_t,x_t,p_t)$.

\begin{rem}\label{rem:benchmark} Our benchmark is weaker than what is typically assumed in literature that knows the 
realizations of  $v_t, x_t(b), p_t(b), t=1,\dots, T$ for all $b$ at time $1$ itself and makes its bids $b_t$ while satisfying the ROSC. In particular, the 
benchmark's reward is 
\begin{align}\nn
 & \quad {\bar V}(T)=  \bbE\left\{ \max_{b_t}\ \sum_{t=1}^T v_t x_t(b_t)\right\},\\\label{eq:optval}
\text{subj. to} &  \quad \bbE\left\{\sum_{t=1}^Tp_t(b_t)\right\} \le   \bbE\left\{\ \sum_{t=1}^T v_t x_t(b_t)\right\}.
\end{align}
We are considering a weaker benchmark for two reasons; i) it is more fair comparison for an online algorithm and ii) our primary interest is in deriving  lower bounds, which are stronger with  with weaker benchmarks.
\end{rem}

To characterize the performance of any online learning algorithm $\cA$, we define its  regret as 
\begin{equation}\label{defn:regret}
\cR_\cA(T) = \max_{\cD} \left(V_{\opt}(T) - V_{\cA}(T)\right) 
\end{equation}
where $\cD$ is the distribution for $(v_t,x_t,p_t)$.
We have defined ROSC as a constraint in expectation. With abuse of notation, its {\bf sample path} counterpart is defined as {\bf ROSC-S}.
\begin{definition}\label{defn:ccv}
Let $\text{CV}(t ) = p_t(b_t) - v_t x_t(b_t)$ be the contribution to ROSC-S  at slot $t$.
Let $\text{CCV}(\tau) = \sum_{t=1}^\tau \text{CV}(t)$ denote the cumulative contribution to ROSC-S till slot $\tau$. 
Let $\text{Margin}(\tau) = \max\{0, - \text{CCV}(\tau)\}$. For ROSC to be satisfied, $\bbE\{\text{CCV}(T)\}\le 0$.
\end{definition}

{\bf For deriving our lower bounds} we will consider the following special class of the allocation and payment functions $x_t(b),p_t(b)$ that are of threshold type, denoted by $x_t^\theta, p_t^\theta$, defined as follows.
\begin{align} \label{def:thresholdfunc}
x_t(b) =
&\left\{
\begin{aligned}
1 & \ \ \text{for}\ \ \ b \ge \theta,\\
0 &\ \  \text{otherwise,}
\end{aligned}\right.
& 
p_t^\theta(b) = 
\left\{\begin{aligned}
 \theta &\ \  \text{for}\ b \ge \theta,\\
 0 &\ \  \text{otherwise,}
\end{aligned}\right. 
\end{align}
where $\theta\le 1$ without loss of generality (WLOG) since $v_t\le 1$.
When we consider allocation function of type \eqref{def:thresholdfunc}, succinctly we write the input as $(v_t, x_t^\theta)$, where $(v_t, \theta_t)$ is i.i.d. generated from an unknown distribution, and the payment function is \eqref{def:thresholdfunc}.

Note that the threshold allocation and payment functions \eqref{def:thresholdfunc} satisfy the Myerson's
condition \cite{myerson1981optimal}, 
\begin{equation}\label{eq:myersontruthfulcond}
p_t(b) = b x_t(b) - \int_0^b x_t(u) du,
\end{equation}
that is satisfied by large classes of auction, e.g. second price auction. Thus, our derived lower bounds are valid also for structured inputs and for not inputs chosen absolutely adversarially.

 \begin{rem} Myerson's
condition is known to evoke truthful bids in a usual auction where the utility is simply $v_t-p_t(b_t)$ without any constraints. 
\end{rem}

\begin{definition}\label{defn:slotwin}
For allocation functions of type $x_t^{\theta_t}$ \eqref{def:thresholdfunc}, 
an algorithm is defined to {\bf win} a slot if its bid $b_t$ is at least as much as $\theta_t$. A slot that is not won is called {\bf lost}. Let $\b1_{\cA}(t)=1$ if algorithm $\cA$ wins the slot $t$, and is zero otherwise.
\end{definition}
With allocation functions of type \eqref{def:thresholdfunc}, the contribution to ROSC-S comes only from slots that are {\it won}, i.e., when  $b_t\ge \theta_t$, and is equal to $v_t - \theta_t$, which can be both positive and negative. To satisfy the ROSC, the expected positive contribution has to be used judiciously to win slots where the expected contribution is negative.
To illustrate this concretely, consider the following example.
\begin{example}\label{exm:1}
Let $v_t=v=1/2$ for all $t$. Moreover, let the allocation function be of type  \eqref{def:thresholdfunc}, where  $\theta_t \in \{0.4, 0.6, 0.7\}$ with equal probability in any slot $t$. The positive contribution ROSC-S comes from slots where $\theta_t=0.4$ of amount $v-\theta_t=0.1$.
In expectation, there are $T/3$ slots with  $\theta_t=0.4$, thus the total expected positive contribution to ROSC is $0.1 T/3$ which can be 
used by any algorithm to bid for slots when $\theta_t\in \{0.6, 0.7\} > v$, where it can either get a negative per-slot contribution of $0.1$ or $0.2$ towards ROSC-S. To satisfy the ROSC, any algorithm has to win slots such that the sum of the expected negative contribution to ROSC is at least $-0.1 T/3$.
\end{example}
\section{WarmUp}\label{sec:warmup}
To better understand the problem, we first consider the simplest algorithm $\cA_s$ that always bids $b_t=v_t$. From \eqref{eq:paymentcondition}, it follows that $\cA_s$ satisfies the ROSC. Thus, the 
only question is what is the regret of $\cA_s$?

\begin{lemma} $\cR_{\cA_s}(T) =\Omega(T)$ regret even when $v_t=v\  \forall  \ t$.
\end{lemma}
\begin{proof}
Let $v_t = 1/2$ for any $t$. For any slot $t$, let  the allocation function be $x_t^{\theta_t}$ and $\theta_t = v_t \pm \epsilon$ with equal probability with $0 < \epsilon < 1/2$. Then $\cA_s$ that bids $b_t=v_t$ gets a valuation of $v_t=1/2$ with probability $1/2$ whenever $\theta_t=v_t-\epsilon$ and a valuation of $0$ with probability $1/2$ whenever $\theta_t=v_t+\epsilon$. Thus the expected accrued valuation of $\cA_s$ is $\frac{T}{2} \cdot \frac{1}{2}$.

$\opt$ on the other hand can choose to bid $b_t^\opt=\theta_t, \ \forall \ t$,  and accrues a valuation of $1/2$ on each slot, while satisfying the ROSC, since $p_t^{\theta_t}(b^\opt_t) - v_t x_t^{\theta_t}(b^\opt_t) = \pm \epsilon$ with equal probability.
Thus, $$\cR_{\cA_s}(T) =  \frac{T}{2} - \frac{T}{4} \ge T/4.$$
\end{proof}

We next consider the question of whether a more intelligent algorithm can do better than $\cA_s$ and answer it in the negative.
\section{Lower Bound on Regret of Any Online Algorithm that strictly satisfies the ROSC}\label{sec:gen}
\begin{theorem}\label{thm:lbunivGenV}
$\cR_\cA  = \Omega(T)$ for any $\cA$ that solves \eqref{eq:prob} while strictly satisfying ROSC.
\end{theorem}
Theorem \ref{thm:lbunivGenV} is a {\bf major negative result} and precludes the existence of any online algorithm with sub-linear regret while 
strictly satisfying the ROSC.
The proof of Theorem \ref{thm:lbunivGenV} is provided in Appendix \ref{app:lbnotequal}, 
where the main idea is to construct two inputs 
with small K-L divergence (making them difficult to distinguish for any $\cA$) but for which the behaviour of $\opt$ is very different to ensure that the 
ROSC is satisfied. Interestingly, the lower bound is derived when $v_t$ takes only two possible values $\sfv_1, \sfv_2$ such that that $\sfv_2/\sfv_1<2$. Then $\cA$'s limitation in not knowing the input beforehand is utilized to derive the result. 

Moreover, Theorem \ref{thm:lbunivGenV} is derived by using allocation functions of the type \eqref{def:thresholdfunc} that also satisfy the Myerson's condition. Thus, Theorem \ref{thm:lbunivGenV} is also surprising since the ROSC is a constraint in expectation and there is a lot of structure to the problem, e.g., the input is i.i.d., and the allocation and the payment functions are structured. 

Trivially, Algorithm $\cA_s$ that always bids $b_t=v_t$ has regret $O(T)$. Thus, we get the following result.
\begin{corollary} Algorithm $\cA_s$ that always bids $b_t=v_t$ and exactly satisfies the ROSC is order-wise optimal.
\end{corollary}

\begin{rem} Note that Theorem \ref{thm:lbunivGenV} sheds no light on the refined question: what is the smallest regret possible when ROSC is allowed to be violated by $O(T^\alpha)$ for $0< \alpha<1$.
\end{rem}

 
 In light of Theorem \ref{thm:lbunivGenV}'s negative result,  instead of regret, the more meaningful performance metric for studying Problem \eqref{eq:prob} is the competitive ratio \cite{vaze2023online} defined as follows. 
\begin{definition}\label{defn:compratio} For any $\cA$, its competitive ratio
  $\mu_{\cA} = \min_{\cD} \frac{ \bbE\left\{\ \sum_{t=1}^T v_t x_t(b^\cA_t)\right\} }{ \bbE\left\{\ \sum_{t=1}^T v_t x_t(b^\opt_t)\right\} },$
  where  $\cA$ and $\opt$ satisfy ROSC, and $\cD$ is the unknown distribution.
\end{definition} The objective is to derive an algorithm with largest competitive ratio as possible, which we do in Section \ref{sec:competitive}.

Next, we continue our lower bound expedition and consider Problem \ref{eq:prob} in the repeated identical auction setting where $v_t=v$ for all $t$, where the goal is only to maximize the number of slots won by an algorithm while strictly satisfying the ROSC.
\section{Repeated Identical Auction Setting $v_t=v  >0\ \forall \ t.$}
\subsection{Lower Bound on Regret for Any Online Algorithm $\cA$}
We derive a general result connecting the regret and the $\text{CCV}$ for any online algorithm $\cA$ that solves \eqref{eq:prob} when $v_t=v$ for all $t$, as follows.

\begin{theorem}\label{thm:lbuniv}
Even when $v_t=v, \ \forall \ t$, for any  $\cA$ that solves \eqref{eq:prob}, we have that 
  $\cR_\cA(T) + \bbE\{\text{CCV}_\cA(T)\} = \Omega(\sqrt{T}).$
\end{theorem}

As a corollary of Theorem \ref{thm:lbuniv}, we get that the following  {\bf main result} of this section.
\begin{lemma}\label{lem:lbuniv}
Even when $v_t=v, \ \forall \ t$, $\cR_\cA(T)= \Omega(\sqrt{T})$ for any $\cA$ that solves \eqref{eq:prob} and strictly satisfies the ROSC.
\end{lemma}
The proof of Theorem \ref{thm:lbuniv} is provided in Appendix \ref{app:lbequal}, where the main idea is similar to that of Theorem \ref{thm:lbunivGenV}
 of constructing two inputs 
with small K-L divergence (making them difficult to distinguish for any $\cA$) but for which the behaviour of $\opt$ is very different to ensure that the 
ROSC is satisfied. Compared to Theorem \ref{thm:lbunivGenV}
 there is less freedom since $v_t=v$ for all $t$, and hence we get a weaker lower bound. 

Lemma \ref{lem:lbuniv} exposes the inherent difficulty in solving Problem \ref{eq:prob}, that is even if $v_t=v$ is known, the problem remains 
challenging, and either the regret or the ROSC violation has to scale as $\Omega(\sqrt{T})$. 

Currently, the best known algorithm \cite{Feng} for solving Problem \eqref{eq:prob} when Myerson's condition \eqref{eq:myersontruthfulcond} is satisfied has both regret and ROSC violation guarantee of $O(\sqrt{T})$. Thus, Theorem \ref{thm:lbuniv} does not imply a lower bound of $\Omega(\sqrt{T})$ on the regret of algorithm \cite{Feng}. In the next section, we separately show that the regret of algorithm \cite{Feng} is $\Omega(\sqrt{T})$ irrespective of its constraint violation even in the repeated identical auction setting. This is accomplished by connecting the regret of algorithm \cite{Feng} to the hitting probability of a reflected random walk. 

\section{Lower Bound on Regret for algorithm \cite{Feng} }

The primal-dual based algorithm $\cA_{pd}$ constructed in \cite{Feng}  for minimizing the regret \eqref{defn:regret}, chooses
$$b_t = v_t+\frac{v_t}{\lambda_t},$$ where 
$$\lambda_t = \exp\left(\alpha \ \text{CCV}(t-1)\right)$$ with $\lambda_1=1$ and $\alpha=1/\sqrt{T}$.
From \cite{Feng}, we have $\cR_{\cA_{pd}}(T) =O(\sqrt{T})$ and ROSC-S of $\text{CCV}(T) = O(\sqrt{T\log T})$. 
Next, we show that $\cR_{\cA_{pd}}(T) =\Omega(\sqrt{T})$.
\begin{lemma}\label{lem:lbApd} $\cR_{\cA_{pd}}(T) =\Omega(\sqrt{T})$ even if $v_t=v \ \forall \ t$ and the Myerson's condition \eqref{eq:myersontruthfulcond} is satisfied.
\end{lemma}
\begin{proof}
{\bf Input:} Let $v_t=v=1/2 \ \forall \ t$. Let the allocation function be $x_t^{\theta_t}$ \eqref{def:thresholdfunc} in any slot, where $\theta_t = \{0,1\}$ with equal probability which satisfies the Myerson's condition \eqref{eq:myersontruthfulcond}. Since $\theta_t$ is symmetric around $v_t=1/2$,  $\opt$ can bid $b_t=\theta_t$ for each slot $t$, and win every slot resulting in its total accrued valuation of $vT = T/2$, while clearly satisfying  the ROSC.

Since $v_t=1/2 \ \forall \ t$, the  total accrued valuation of algorithm $\cA_{pd}$ is 
$$\sum_{t=1}^T \b1_{\cA_{pd}}(t) v = \sum_{t=1}^T \b1_{\cA_{pd}}(t) \frac{1}{2}.$$
Note that because of the choice of $v_t$ and $\theta_t$, $\text{CV}(t) = p_t^{\theta_t}(b_t) - v_t x_t^{\theta_t}(b_t)$ takes only two values $\left\{-\frac{1}{2}, \frac{1}{2}\right\}$ 
for a slot that is {\bf won} by $\cA_{pd}$. For a slot that is {\bf lost}, $\text{CV}(t)=0$.

By the choice of the algorithm $\cA_{pd}$, whenever $\text{CCV}(t-1) = 1/2$, $\lambda_t= \exp\left(\alpha \ \text{CCV}(t-1)\right)\ge 1$ and its bid $b_t = v_t + \frac{v_t}{\lambda_t}< 1$ and $\cA_{pd}$ wins that slot $t$ only if $\theta_t=0$.
Therefore the CCV process for $\cA_{pd}$ evolves as follows, 
\begin{equation}\label{defn:CCVprocessFeng}
\text{CCV}(t)=\begin{cases}\text{CCV}(t-1) \pm 1/2 \ \quad   \text{w.p. $\{1/2, 1/2\}$ if} \  
\text{CCV}(t-1) < 1/2, \\
\text{CCV}(t-1) -1/2 \ \ \quad \text{if}\ \theta_t=0 \ \ \ \text{and} \
\text{CCV}(t-1) =1/2,
\\
\text{CCV}(t-1)\ \quad \quad \quad \quad  \text{if}\ \theta_t=1\ \ \   \text{and} \  
\text{CCV}(t-1) = 1/2, \end{cases}
\end{equation}
where the first equation for $\text{CCV}(t-1) < 1/2$ in \eqref{defn:CCVprocessFeng} follows for the following reason. Recall that the per-slot $\text{CV}(t) \in \{-1/2, 0, 1/2\}$. Thus $\text{CCV}(t)$, only takes values that are integral multiples of $1/2$. Thus, when $\text{CCV}(t-1)<1/2$, it is in fact $\text{CCV}(t-1)\le 0$ and hence $\lambda_t \le 1$ and $\cA_{pd}$ wins every slot and the increment $\text{CV}(t)$ to $\text{CCV}(t)$ takes only two values $\{-1/2, 1/2\}$ with equal probability corresponding to $\theta_t=0$ and $1$, respectively.

Thus, the $\text{CCV}(t)$ process is a simple reflected (backwards at $1/2$) random walk with increments $\pm 1/2$ with equal probability. Using classical results on reflected random walks we proceed further.

\begin{definition}\label{defn:randomwalk}\cite{port1965}
Consider a simple random walk process $\cW$ with i.i.d. increments  $X_i$ (a random variable) having $\bbE\{X_1\}=0$ and $\bbE\{X_1^2\}<\infty$  reflected at zero, i.e. $\cW_n = \max\{\cW_{n-1}+X_n,0\}$, where 
$\cW_0=0$.
 Let $\bbP_n(x,y)$ be the probability that $\cW_n=y$ given $\cW_0=x$.
\end{definition}

\begin{prop}\label{prop:classicalrandomwalkresult1}\cite{port1965} 
For Definition \ref{defn:randomwalk},  
$$\lim_{n\rightarrow \infty} \bbP_n(0,0)\sqrt{n} = c,$$ for constant $c>0$. Moreover, $$\lim_{n\rightarrow \infty} \frac{\bbP_{n+1}(x,0)}{\bbP_n(0,0)} =1, \ \forall \ x. $$
\end{prop}


Using the two parts of Proposition \ref{prop:classicalrandomwalkresult1} for the CCV process \eqref{defn:CCVprocessFeng}, and noting that \eqref{defn:CCVprocessFeng} is reflected at $1/2$, we have 
$\bbP(\text{CCV}(t) = 1/2)\ge c/ \sqrt{t}$ for $t$ large.
Whenever, $\text{CCV}(t) =1/2$,  $\cA_{pd}$'s bid is $<1$ and it wins a slot $t$ only if $\theta_t=0$, which happens with probability $1/2$.
Therefore, using linearity of expectation, we get that the expected number of  slots {\bf lost} by $\cA_{pd}$ is 
$\ge \frac{1}{2} \sum_{t=T/2}^T \frac{c}{\sqrt{t}} = \Omega(\sqrt{T}),$
where we are accounting only for $t\ge T/2$ since Proposition \ref{prop:classicalrandomwalkresult1} applies for large $t$. 
Recall that $\opt$ wins all the slots, thus the regret of $\cA_{pd}$ is at least as much as its number of lost slots.
\end{proof}

As far as we know there is no known algorithm even in the repeated auction setting $v_t=v \ \forall \ t $  for Problem \ref{eq:prob} that strictly satisfies the ROSC and achieves the lower bound of Lemma \ref{lem:lbuniv} unless one considers a trivial setting \cite{Feng} to be described in Remark \ref{rem:degenerate}. 
In the next section, we construct one such algorithm when the allocation and payment function are of the threshold type \eqref{def:thresholdfunc} under some mild assumptions.  
%

\section{Algorithm with regret $O(\sqrt{T \log T})$ in the Repeated Identical Auction strictly satisfying the ROSC}
Note that from Lemma \ref{lem:lbuniv}, we know that even the constant $v$ case ($v_t=v >0 \ \forall \ t$) is not trivial for any $\cA$ to solve \eqref{eq:prob} that strictly satisfies the ROSC. 
To belabour the point, consider a simple algorithm $\cA_b$ that satisfies the ROSC on a sample path basis, i.e., $\text{CCV}(T)\le 0$, by  bidding $$b_t = v_t+\text{Margin}(t-1),$$ where $\text{Margin}(t-1)$ has been defined in Definition \ref{defn:ccv}.

Using Example \ref{exm:1}, we next show that $\cR_{\cA_b}(T)=\Omega(T)$. Let the allocation function be of type \eqref{def:thresholdfunc}, where $\cA_b$ wins a slot if $b_t \ge \theta_t$. 
Since $v_t$ is a constant, any algorithm is just trying to maximize the number of slots it can win while satisfying the ROSC. 
First we describe the $\opt$. For the input defined in Example \ref{exm:1},  slots with $\theta_t=0.4$ are won for `free' since $\opt$ always bids $b_t\ge v=1/2$, and provide with a positive contribution towards ROSC-S of $0.1$ in each such slot. 
Since $v_t=v \ \forall \ t$, among slots where $\theta_t>v$, $\opt$ only wins those slots where $\theta_t=0.6$ (negative contribution towards ROSC-S of $0.1$) by bidding $b_t=0.6$ to ensure satisfying the ROSC, since $\theta_t$ takes values $0.4$ and $0.6$ with probability $1/3$. Thus, $b_t^\opt=0.6$ for all $t$, and $V_\opt = T\cdot v \cdot (2/3)$.

  \begin{figure*}
        \includegraphics[scale=.5]{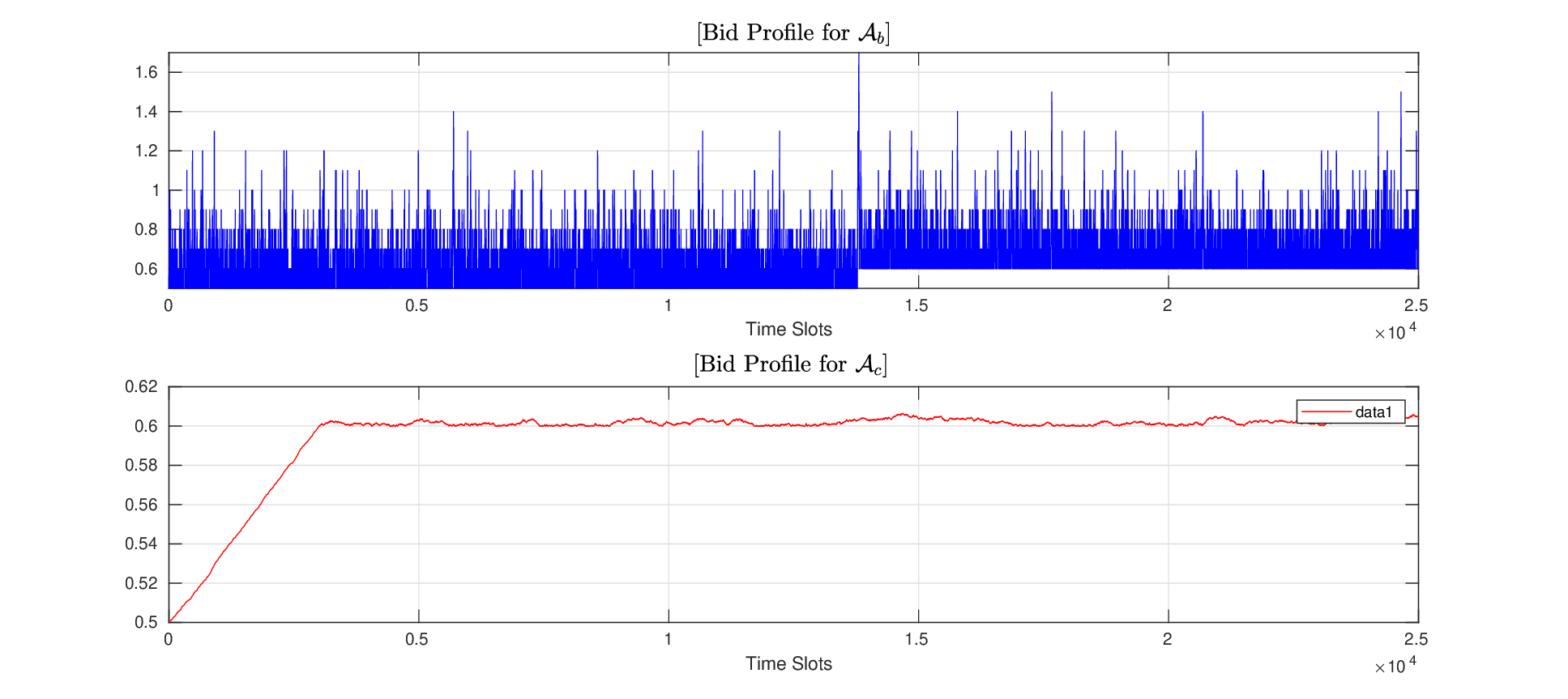}
        \caption{Bid Profiles for algorithm $\cA_b$ and $\cA_c$ for Example \ref{exm:1}.}
        \label{fig:1}
    \end{figure*}

The bid profile of $\cA_b$ is, however, different than $\opt$, as can be seen  from Fig. \ref{fig:1}, where it wins slots with both $\theta=0.6$ and $\theta=0.7$ because of large variability in its bid values $b_t$. Winning a single slot with $\theta=0.7$ precludes winning two slots with $\theta=0.6$, and since the objective is to maximize the total number of won slots, $\cA_b$ falls behind $\opt$ and has linear regret with respect to $\opt$ since it wins slots having $\theta=0.7$ with non-zero probability. This can be shown rigorously by using the same results on random walks as done in Lemma \ref{lem:lbApd}.

Thus, an algorithm that hopes to achieve a sub-linear regret has 
to inherently learn the maximum bid $\opt$ will make.
Towards that end, we propose algorithm $\cA_c$ that bids
 \begin{equation}\label{eq:bidprocess}
b_t = v+ \frac{1}{\sqrt{T}}\text{Margin}(t-1)
\end{equation}
 where the Margin process is as defined in Definition \ref{defn:ccv}.
Similar to $\cA_b$, $\cA_c$ also satisfies the ROSC  on a sample path basis.


For allocations functions satisfying \eqref{def:thresholdfunc}, the Margin process  simplifies to 
   \begin{equation}\label{defn:marginprocess}
\text{Margin}(t) = \begin{cases} \text{Margin}(t-1) + (v-\theta_t) & \text{if} \ b_t \ge \theta_t 
\\ 
 \text{Margin}(t-1), & \text{otherwise}.
 \end{cases}
\end{equation}
$\cA_c$ is a conservative algorithm that overbids by an amount exactly equal to its non-negative slack in ROSC-S until that slot scaled down by $\sqrt{T}$. 
Motivation behind $\cA_c$ can be best understood by following the execution of $\cA_c$ again over Example \ref{exm:1}.
$\cA_c$, in contrast to $\opt$, initially does not win any slots with $\theta_t \in \{0.6, 0.7\}$, but once its bid value `converges' (as can be seen in Fig. \ref{fig:1}), wins all slots $t$ with $\theta_t =0.6$ and no slot $t$ with $\theta_t=0.7$. With $\cA_c$, $b_t$ crosses the optimal maximum bid value of $0.6$ of $\opt$ in $O(\sqrt{T})$ time and stays very close to it throughout, thereby avoiding winning any slots with $\theta_t=0.7$. 


In general, $\cA_c$ lets its bid process converge to a value close to the maximum bid that $\opt$ will make, and then operates its bid profile in a narrow band around that value which avoids winning slots with large values of $\theta$ (that $\opt$ avoids), while ensuring that enough slots that $\opt$ wins are also won by it. To achieve this, $\cA_c$ sacrifices a regret of $O(\sqrt{T})$ in the initial phase where its bid is very low because of scaling down by $1/\sqrt{T}$.
Next, we make this intuition formal, and the {\bf main result} of this section is as follows.

\begin{theorem}\label{thm:ubequalcase}
 When $v_t=v >0, \ \forall \ t$, the regret of $\cA_c$ is at most $O(\sqrt{T \log T})$ when allocation functions are of type \eqref{def:thresholdfunc} and Assumptions \ref{assump:drift} and \ref{assump:updown}  holds.
\end{theorem}
\begin{rem} Theorem \ref{thm:ubequalcase} holds against the stronger benchmark \eqref{eq:optval} (Remark \ref{rem:benchmark}) as well with an identical proof.
\end{rem}

Combining Theorem \ref{thm:ubequalcase} with  Lemma \ref{lem:lbuniv}, we conclude that algorithm $\cA_c$ achieves optimal regret up to logarithmic terms. Even though Theorem \ref{thm:ubequalcase} is applicable only for allocation functions of type \eqref{def:thresholdfunc} and under some assumptions, it is an important (first) step towards finding algorithms with optimal regret that strictly satisfies the ROSC. Conceptually, $\cA_c$ is expected to work well for general allocation functions, however, how to generalize 
the regret guarantee is not obvious, and remains a challenging open question.

Full proof of Theorem \ref{thm:ubequalcase} is provided in Appendix \ref{app:thm:ubequalcase}, with the following proof outline.
As assumed, let $v_t=v>0, \ \forall \ t$. 
  We first define a few quantities.
 For fixed $v$, let  $\mu_L = \bbE\{1_{v \ge \theta} \cdot(v-\theta)\}$, while $\mu_R = \bbE\{1_{v< \theta}\cdot (v-\theta)\}$. Note that $\mu_L$ and $\mu_R$ 
 are expected per-slot positive and negative contributions to the ROSC, respectively, if all slots are won, e.g. by bidding $b_t=1 \ \forall \ t$.
 Let $\mu_L + \mu_R \le 0$. The other case is easier to handle since then the per-slot expected contribution to ROSC is non-negative, and an algorithm can win all slots by bidding $b_t=1$ while 
satisfying the ROSC.
 
 \begin{rem}\label{rem:degenerate} Algorithm $\cA_{pd}$ of \cite{Feng} is shown to satisfy the ROSC when $\mu_L + \mu_R > 0$. But in this setting the problem is trivial, since bidding $b_t=1$ always 
 guarantees zero regret while satisfying the ROSC  in the repeated identical auction setting.
 \end{rem}
 
Given that $v_t=v$ for all $t$, the objective is to maximize the total number of won slots while satisfying the ROSC. Essentially, the identity of which slots are won is immaterial. Therefore, for $\opt$, there exists a $\theta^\star$ and $0<\pi^\star\le 1$ such that it bids 
$b_t = \theta_t$ whenever $\theta_t<\theta^\star$ and wins all those slots, while bids 
$b_t=\theta^\star$ with probability $\pi^\star$ when $\theta_t=\theta^\star$ (wins slots with $\theta_t=\theta^\star$ with probability $\pi^\star$) such 
that the ROSC is satisfied. In particular, 
\begin{equation}\label{defn:thetastar}
\pi^\star \bbE\{1_{v= \theta^\star}\cdot (v-\theta^\star)\} + \bbE\{1_{v< \theta^\star}\cdot (v-\theta^\star)\}+ \mu_L = 0.
\end{equation}
 
Let $\underline {\mu}_R^{\theta^\star} = 
\bbE\{\b1_{v< \theta}\cdot (v-\theta) | \theta<\theta^\star\}$ and $\bar {\mu}_R^{\theta^\star} = 
\bbE\{\b1_{v< \theta}\cdot (v-\theta) | \theta \ge \theta^\star\}$.  
  \begin{definition}\label{defn:mindrift}
For $\cA_c$, the increments (positive jumps) to the Margin process \eqref{defn:marginprocess} have mean $\mu_L$, while the mean of the decrements (negative jumps) to the Margin process \eqref{defn:marginprocess} is  $\underline {\mu}_R^{\theta^\star}$ (a negative quantity) until $b_t<\theta^\star$. Thus, the $\text{Margin}(t)$ process \eqref{defn:marginprocess} has  drift $\Delta = \mu_L +\underline {\mu}_R^{\theta^\star}$ until $b_t<\theta^\star$.
Following \eqref{defn:thetastar}, $\Delta>0$. 
\end{definition}
 \begin{assumption}\label{assump:drift} $\Delta$ does not depend on $T$.
 \end{assumption}
  \begin{assumption}\label{assump:updown} $|\bar {\mu}_R^{\theta^\star}| \le |\underline {\mu}_R^{\theta^\star}|$.
 \end{assumption}
 Assumption \ref{assump:drift} implies that the positive drift (function of unknown distribution $\cD$) of the Margin process of $\cA_c$ does not depend on $T$, which is necessary. For example,  if $\Delta= 1/T$, it will take $\Omega(T)$ time for algorithm $\cA_c$'s bid to cross any constant level, making its regret $\Omega(T)$.

Assumption \ref{assump:updown} implies that the negative drift to $\text{Margin}(t)$ process \eqref{defn:marginprocess} obtained by winning slots with $\theta_t > v$ is weaker after the bid process crosses $\theta^{\star}$ compared to before crossing $\theta^{\star}$. Thus, the excursions of the Margin process around $\theta^\star$ are `well behaved' and we can use the result from \cite{Koksal}.
 It is worth noting that Assumption \ref{assump:updown} does not make the problem trivial, and is needed for technical reasons in the proof.


  \begin{definition}
Let $\tau_{\min}= \min\{t: b_t \ge \theta^\star\}$ be the earliest time at which $b_t$ process \eqref{eq:bidprocess} crosses $\theta^\star$. 
\end{definition}
\begin{definition}\label{defn:slotindex}Let the set of all $T$ slots be denoted by $\cT$. For $t\ge  \tau_{\min} $ the subset of slots where $\theta=\theta^\star$ be denoted by $\cT_{\theta^\star}$, while where $\theta> \theta^\star$ by $\cT^+_{\theta^\star}$ and $\theta< \theta^\star$ by $\cT^{-}_{\theta^\star}$. 
\end{definition}

Using Definition \ref{defn:slotindex} we can write the regret for $\cA_c$ as 
\begin{equation}\nn
\cR_{\cA_c}(T)= \cR_{\cA_c}([1:\tau_{\min}]) +   \cR_{\cA_c}( \cT_{\theta^\star})  +  \cR_{\cA_c}(\cT^+_{\theta^\star})+ \cR_{\cA_c}(\cT^{-}_{\theta^\star}). 
\end{equation}
From definition \eqref{defn:thetastar}, $\opt$ does not win any slots of $\cT^+_{\theta^\star}$, thus, 
$\cR_{\cA_c}(\cT^+_{\theta^\star})\le 0$.
Under Assumption \ref{assump:drift}, the rest of the {\bf proof of Theorem \ref{thm:ubequalcase} is divided in three parts}, i) we show that $\bbE\{\tau_{\min}\} = O(\sqrt{T})$ which implies that $\cR_{\cA}([1:\tau_{\min}])=O(\sqrt{T})$, ii) $\cR_{\cA_c}(\cT_{\theta^\star})\le 0$  by showing that $\cA_c$ and $\opt$ win equal number of slots in expectation when $\theta = \theta^\star$, and finally iii) $\cR_{\cA_c}(\cT^{-}_{\theta^\star}) = O(\sqrt{T \log T})$ using a result from \cite{Koksal} stated in Lemma \ref{lem:RAstar} that shows that bid process \eqref{eq:bidprocess} concentrates very close to $\theta^\star$, and $\cA_c$ wins any slots having $\theta<\theta^\star$ with probability 
$\Omega(1-1/\sqrt{T})$. Putting all of this together implies the result.




We next move on to the more challenging and interesting case when $v_t\ne v, \ \forall \ t$ and derive a $1/2$-competitive algorithm.

\section{$1/2$-Competitive Algorithm for Problem \ref{eq:prob}}\label{sec:competitive}

In Section \ref{sec:gen}, we showed that regret of any online algorithm for solving \eqref{eq:prob} is $\Omega(T)$ while strictly satisfying the 
ROSC. In this section, hence we consider a weaker multiplicative guarantee of competitive ratio (Definition \ref{defn:compratio}), and present an algorithm that can achieve a competitive ratio of close to $1/2$ while satisfying the ROSC with high probability.

In this section, for notational simplicity, we consider the discrete version of Problem \ref{eq:prob}, where in any slot $t$, $v_t$ takes $K$ (arbitrary) possible values $\sfv_1 > \dots > \sfv_K, \sfv_1\le 1$ with probabilities $p_1, \dots, p_K$ that are independent of $T$. Note that $K$, $\sfv_1 > \dots > \sfv_K$ and $p_1, \dots, p_K$ are unknown to begin with. We consider allocation and payment functions 
of type \eqref{def:thresholdfunc}. The results generalize in a straightforward manner for general allocation and payment functions unlike Theorem \ref{thm:ubequalcase} which is specific to allocation and payment functions 
of type \eqref{def:thresholdfunc}.

Let $\cD$, the distribution for $(v_t,x_t,p_t)$ be known. Consequently, 
let the expected per-slot positive and negative  contribution to ROSC when $v_t=\sfv_k$ be 
$\ell_k = \bbE\{\sfv_k - \theta_t|v_t=\sfv_k,  \theta_t\le \sfv_k\}$ and $r_k = \bbE\{ \theta_t-\sfv_k |v_t=\sfv_k,  \theta_t> \sfv_k\}$, respectively. Moreover, let  
$\overleftarrow{w}_k = \bbP( \theta_t\le v_t|v_t= \sfv_k)$ and $\overrightarrow{w}_k = \bbP( \theta_t > v_t|v_t= \sfv_k)$. Thus, $\sum_{i=1}^k p_i\overleftarrow{w}_i \ell_i$ is the total expected positive contribution to ROSC which can be used to win slots where $\theta_t > v_t$.



Since $\sfv_1\le 1$ and $\theta_t \le 1$ for all $t$, let an algorithm bid $b_t = 1$ when $v_t=\sfv_k$ with probability $q_k$. 
Doing so, if $\theta_t \le v_t$, then bidding $b_t = 1$ is same as bidding $b_t = v_t$. In case $\theta_t > v_t$, then by bidding $b_t=1$, 
the expected per-slot negative contribution to the ROSC is simply $r_k$. 
  Then the benchmark $\opt$'s solution (which is assumed to know $\cD$) to problem \eqref{eq:prob} can be recast as the following LP, 
\begin{align}\nonumber
\max_{  \  0\le q_k \le 1} & \quad \sum_{k=1}^K p_k\overleftarrow{w}_k \sfv_k + \sum_{k=1}^K p_k\overrightarrow{w}_k \sfv_k  q_k \\ \label{eq:optprob}
\text{subj. to} & \quad  \sum_{k=1}^K p_k\overrightarrow{w}_k r_k q_k + \sum_{k=1}^K p_k \overleftarrow{w}_k \ell_k \ge 0, \ \forall \ k,
\end{align}
where the constraint in \eqref{eq:optprob} exactly corresponds to the ROSC. 
Let $\cP_{\text{light}} =\{K, p_k, \overleftarrow{w}_k, \overrightarrow{w}_k, r_k, \ell_k\}$.
 Assuming the knowledge of $\cP_{\text{light}}$, \eqref{eq:optprob} is similar to the fractional knapsack problem \cite{vaze2023online} and the optimal solution has the following structure. 

\begin{lemma}\label{lem:order}
For \eqref{eq:optprob},  $\exists$ a threshold $\sigma^\star$, such that for indices $k$ such that for $\frac{ \sfv_k}{r_k} > \sigma^\star$, $q_k^\star = 1$, and for $\frac{ \sfv_k}{r_k} < \sigma^\star$, $q_k^\star = 0$. The only fractional solution $q_k^\star$ is for index $k$ such that $\frac{ \sfv_k}{r_k}=\sigma^\star$.
\end{lemma}
 Assuming the knowledge of $\cP_{\text{light}}$
Lemma \ref{lem:order} implies an optimal algorithm to find the exact solution of \eqref{eq:optprob}. Solve \eqref{eq:optprob} to find $\sigma^\star$. 
Bid $b_t=1$ with probability $1$ for as many $k$'s in order for which $\frac{\sfv_k}{r_k} > \sigma^\star$, bid $b_t=1$ with probability $q<1$ (so as to make the constraint tight in \eqref{eq:optprob}) for index $k$ for which
 $\frac{\sfv_k}{r_k} = \sigma^\star$. For all other $k$'s,  bid $b_t=\sfv_k$, i.e. win such a slot only if $\theta_t \le \sfv_k$ and do not spend any positive accrued contribution to the ROSC-S for winning them.

The quantities involved in \eqref{eq:optprob},
$\cP_{\text{light}} =\{K, p_k, \overleftarrow{w}_k, \overrightarrow{w}_k, r_k, \ell_k\}$, except $K$, are all just expected values and can be easily estimated.
In reality,  $\cP_{\text{light}}$ is unknown, algorithm \textsf{Learn-Alg} (with pseudo-code in Algorithm \ref{alg:light}) first learns the parameters of $\cP_{\text{light}}$ by dedicating the first half of the time horizon $T$, and then applies the solution implied by Lemma \ref{lem:order}.   

\begin{algorithm}
\caption{\textsf{Learn-Alg}}\label{alg:light}
\begin{algorithmic}[1]
\State Begin {\bf Learning Phase}
\State Dedicate slots $t=1, \dots, T/2$ for learning parameters of $\cP_{\text{light}}$ by bidding $b_t=v_t$ and using sample mean estimator ${\hat X}=\frac{1}{n}\sum_{i=1}^n X_i$ for each quantity, ${\hat K} = \#$ distinct values of $v_t$'s observed in time period $[1,T/2]$.
\State Arrange all items in decreasing order of $\frac{ \sfv_k}{r_k} $ for $k=1,\dots, {\hat K}$, and  find $\sigma^\star$ following Lemma \ref{lem:order} with $K={\hat K}$
\State Begin {\bf Decision Phase}
\State For any slot $t\ge T/2$
\State $\ \ $ If $\frac{ \sfv_k}{r_k} > \sigma^\star \quad\quad\  $ Bid $b_t = 1$ 
\State $\quad$Else If $\frac{ \sfv_k}{r_k} = \sigma^\star$ Bid $b_t = 1$ with probability $q_k$:  $q_k$ makes the equality in \eqref{eq:optprob} tight
\State $\quad$Else If $\frac{ \sfv_k}{r_k} < \sigma^\star \quad$ Bid $b_t = v_t$ 
\State End
\end{algorithmic}
\end{algorithm}
\vspace{-0.2in}
\begin{rem}
As defined in Algorithm \ref{alg:light}, ${\hat K}$ is the estimated number of distinct values 
that $v_t$ takes.
Since $p_1, \dots, p_K$ do not depend on $T$, the probability that ${\hat K}\ne K$ is $o(1/T)$. 
\end{rem}

\begin{theorem}\label{proof:learn1}
The competitive ratio (Definition \ref{defn:compratio})
  $$\mu_{\textsf{Learn-Alg}} \ge 1/2 -(1-O(K\log(1/\delta)/\sqrt{T}))$$ and \textsf{Learn-Alg} satisfies the ROSC with  probability $$1-(1-O(K\log(1/\delta)/\sqrt{T}))$$. 
\end{theorem}
\begin{rem}
Choosing the length of the learning phase appropriately, we can tradeoff the competitive ratio with the probability with which the ROSC is satisfied.
\end{rem}
\section{Conclusions}
In this paper, we considered an important online optimization problem, single auto-bidding  under strict ROSC, and 
characterized the performance of any online algorithm. Compared to prior work which allowed ROSC violation, our focus was on enforcing strict ROSC. For the case when the valuation can be different for different slots, we showed an impossibility result that no online algorithm can achieve 
a sub-linear regret, which is a bit surprising. Having closed the door on obtaining a sub-linear regret, we consequently considered the weaker metric of 
competitive ratio and showed that a simple algorithm can achieve a competitive ratio close to one-half. With strict ROSC, we showed that the problem is non-trivial
even if the valuation remains constant across time slots. In particular, we showed that the sum of the regret and the constraint violation scales as square-root of the time horizon, and derived an algorithm for threshold type allocation functions that achieves the regret bound up to logarithmic factors while satisfying the strict ROSC constraint.

\bibliographystyle{elsarticle-num}
\bibliography{refs}

\newpage
\section{Proof of Theorem \ref{thm:lbunivGenV}}\label{app:lbnotequal}
\begin{proof}
Let $v_t$ take two values $\{\sfv_1, \sfv_2\} = \{0.5, 0.9\}$ with equal probability for all $t$.  The allocation functions are $x_t^{\theta_t}$ (of type \eqref{def:thresholdfunc}) where $\theta_t$ are distributed as follows.

Input 1: If for slot $t$, $v_t=\sfv_1$, then $\theta_t \in \{v_t\pm \epsilon\}$ with equal probability, where $0< \epsilon$ small. Moreover, if  for slot $t$,  $v_t=\sfv_2$, then 
$\theta_t = v_t+\sfa+ X$ where $X \in \{0,\sfb\}$ with probability $r, 1-r$, respectively,  and $a + \bbE\{X\} = \sfa+(1-r)\sfb = \epsilon$ with $\sfa,\sfb \ge 0,  \sfv_2+\sfb \le 1$ and  $\sfa>\frac{2\epsilon}{3}$, $r<1-r$, i.e., $r< \frac{1}{2}$.

Input 2: If for slot $t$, $v_t=\sfv_1$, then $\theta_t \in \{\sfv_1\pm \epsilon\}$ with equal probability and if for slot $t$ $v_t=\sfv_2$, then 
$\theta_t = v_t+\sfa+ X'$ where $X'  \in \{0,\sfb\}$ with probability $r-\delta , 1-r + \delta $, respectively with $r-\delta <1-r+\delta$. Let the choice of $\sfa,\sfb,\delta$ be such that $\sfa + \bbE\{X'\} = 4\epsilon$.
Essentially, $\delta=\Theta(\epsilon)$.

See Fig. \ref{fig:inputgen} for a pictorial description where quantities with blue and red color correspond to the values of $v_t$ and $\theta_t$ that appear together.
\begin{center}
\begin{figure}
\includegraphics[width=10cm,keepaspectratio,angle=0]{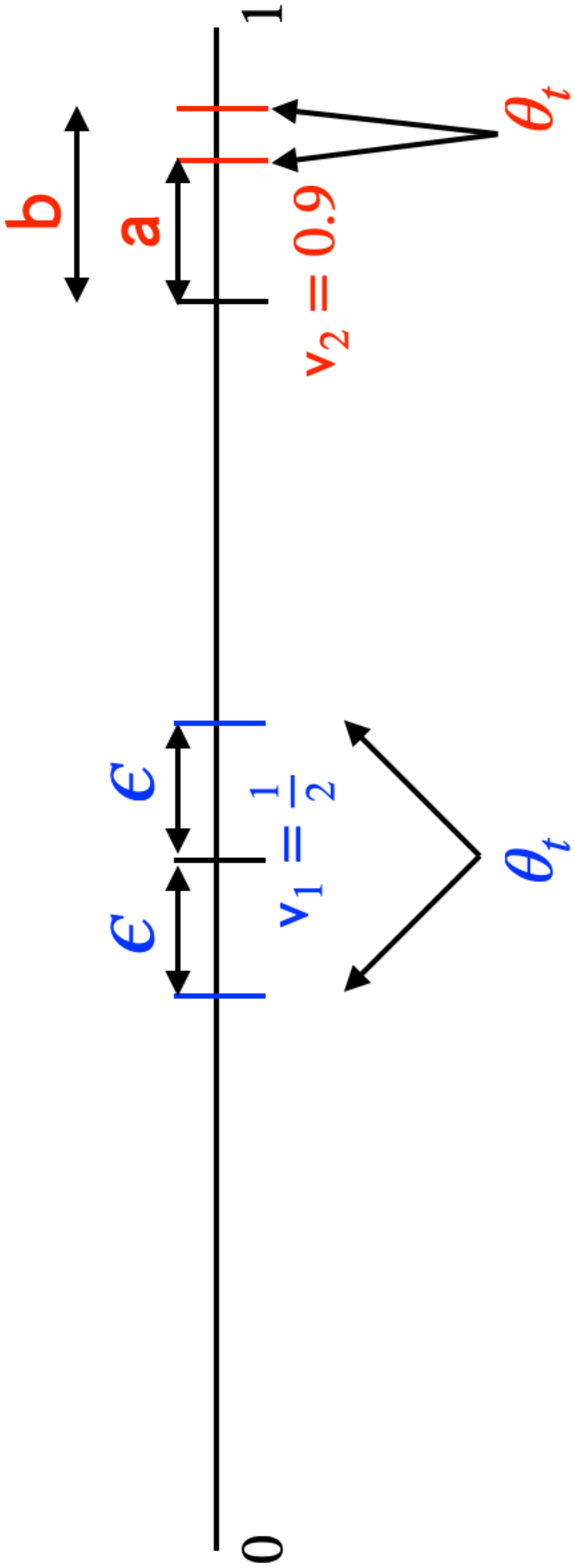}
\caption{Input used to prove Theorem \ref{thm:lbunivGenV}.}
\label{fig:inputgen}
\end{figure}
\end{center}

Note that with both Input 1 and  2, the only case where $\theta_t< v_t$ is when $v_t=\sfv_1$. Thus, given the distribution, the total expected per-slot positive  contribution to ROSC is $\frac{\epsilon}{4}$ with both Input 1 and  2 which can be used 
by $\opt$ or any algorithm to win slots with $\theta_t>v_t$.

{\bf $\opt$'s actions:} By the choice of Input 1, the expected per-slot negative contribution to ROSC by winning a slot with $v_t=\sfv_1$ and $\theta_t=v_t+\epsilon$ is $-\epsilon/4$ while the expected per-slot negative contribution to ROSC by winning a slot with $v_t=\sfv_2$ and $\theta_t=v_t+X$ is $-\epsilon/2$. 
Moreover, the expected value from winning a slot with $v_t=\sfv_1$ and $\theta_t=v_t+\epsilon$ is $\sfv_1/4$ while by winning a slot with $v_t=\sfv_2$ and $\theta_t=v_t+X$ is $\sfv_2/2$.
Thus, since $\sfv_2>\sfv_1$, with Input 1, consider an algorithm $\cB$ that spends all the available 
total expected per-slot positive contribution to ROSC of $\frac{\epsilon}{4}$ in winning the maximum number slots with $v_t=\sfv_2$ while satisfying the ROSC. Thus, $\cB$'s bidding profile on input 1 is:  bid $b_t=v_t$ whenever $v_t=\sfv_1$, and bid $b_t=1$ with probability $1/2$ whenever $v_t=\sfv_2$. 
Thus, $\cB$ satisfies the ROSC exactly, and its expected accrued valuation is $\sfv_1T/4+\sfv_2 T/4$. Since $\opt$ is at least as good as any other algorithm while satisfying the ROSC exactly, $\opt$'s expected accrued valuation is $\ge \sfv_1T/4+\sfv_2 T/4$.


Next, we characterize the structure of $\opt$ with Input 2.
\begin{definition}\label{defn:p3}
Let any algorithm bid $b_t=v_t+\epsilon$ whenever $v_t=\sfv_1$ and $\theta_t=\sfv_1+\epsilon$, with probability $g_1$,  
bid $b_t=v_t+\sfa$ whenever $v_t=\sfv_2$ and $\theta_t=\sfv_2+\sfa$, with probability $g_2$, and bid $b_t=v_t+\sfb$ whenever $v_t=\sfv_2$ and $\theta_t=\sfv_2+\sfb$, with probability $g_3$.
\end{definition}
Following Definition \ref{defn:p3}, the expected accrued valuation per-slot by $\opt$ with Input 2 is
\begin{equation}\label{eq:accvalLBGen}
\max_{g_1, g_2, g_3} \ \ \frac{1}{4}\sfv_1 + \frac{1}{4}g_1 \sfv_1 + g_2 \sfv_2 \frac{1}{2} \left(r-\delta\right) + g_3 \sfv_2 \frac{1}{2} \left(1-r+\delta\right)
\end{equation}
subject to satisfying the ROSC that translates to
\begin{equation}\label{eq:consLBGen}
g_1 \frac{1}{4} \epsilon + g_2 \sfa \frac{1}{2} \left(r-\delta\right)  + g_3 (\sfa+\sfb) \frac{1}{2} \left(1-r+\delta\right) \le \frac{\epsilon}{4}.
\end{equation}
\begin{lemma}\label{lem:accvalLBGen} With Input 2, let $\bp^\star = (g_1^\star, g_2^\star, g_3^\star)$ be the $\opt$'s solution to maximize \eqref{eq:accvalLBGen} satisfying \eqref{eq:consLBGen}. Then $g_3^\star=0$.
\end{lemma}
\begin{proof}
Let $\opt$'s solution be $\bg_1 = (g_1, g_2, g_3)$ to maximize \eqref{eq:accvalLBGen} satisfying \eqref{eq:consLBGen}, where $g_3>0$.  To prove the result, we will perturb 
$\bg_1$ to create a new solution $\bg_1'$ that has $g_3'=0$ and a higher valuation than that of $\bg_1$, as follows.
It is easy to see that for $\opt$, $g_2\ge g_3$, since the expected per-slot negative contribution to ROSC with the two choices which are chosen with probability $g_2$ and $g_3$ are $-\sfa \frac{1}{2} \left(r-\delta\right)$ and $-(\sfa+\sfb) \frac{1}{2} \left(1-r+\delta\right)$, respectively, where $ r-\delta <1-r+\delta$, while the accrued value is $\sfv_2$ in both cases. 
Also for $\opt$, the constraint in \eqref{eq:consLBGen} will be tight. Thus, \eqref{eq:consLBGen} is 
\begin{align}\nonumber
g_1 \frac{1}{4} \epsilon + g_2 \sfa \frac{1}{2} \left(r-\delta\right)  + g_3 (\sfa+\sfb) \frac{1}{2} \left(1-r+\delta\right) &= \frac{\epsilon}{4}, \\ \nonumber
 g_1 \frac{1}{4} \epsilon + (g_2-g_3) \sfa \frac{1}{2} \left(r-\delta\right) + \frac{g_3}{2} \left(\sfa  \left(r-\delta\right)  + (\sfa+\sfb) \left(1-r+\delta\right)\right) &\stackrel{(a)}= \frac{\epsilon}{4}, \\\label{eq:dummy11x1}
g_1 \frac{1}{4} \epsilon + (g_2-g_3) \sfa \frac{1}{2} \left(r-\delta\right) + \frac{g_3}{2} 4 \epsilon &\stackrel{(b)}= \frac{\epsilon}{4},
\end{align}
where $(a)$ follows by adding and subtracting $g_3 \sfa \frac{1}{2} \left(r-\delta\right)$ to the L.H.S. while (b) follows since $ \sfa  \left(r-\delta\right) + (\sfa+\sfb)  \left(1-r+\delta\right) =\sfa + \bbE\{X'\} = 4 \epsilon$.
Compared to $\bg_1 = (g_1, g_2, g_3)$, consider another solution $\bg_1' = (g_1+8g_3, g_2-g_3, 0)$ that also satisfies \eqref{eq:dummy11x1}. Moreover,
the valuation  \eqref{eq:accvalLBGen} with $\bg_1'$ is larger than $\bg_1$, since $r-\delta\le 1$, $(1-r+\delta) \le 1$, and $\sfv_2/\sfv_1=0.9/0.5=1.8$.
Hence, $\bg_1$ cannot be optimal when $g_3>0$.
\end{proof}

To summarize, with Input 2, $\opt$ never bids $b_t >  v_t+\sfa$ whenever $v_t=\sfv_2$.
The following corollary is immediate.
\begin{corollary}\label{cor:regretLB}
 With Input 2, any online algorithm $\cA$ that uses $g_3>\sfc>0$ ($\sfc$ is a constant) has regret of $\Omega(T)$ or $\cA$ violates the ROSC.
\end{corollary}
Winning slots with option i) $v_t=\sfv_2, \theta_t=\sfv_2+\sfa+\sfb$ compared to option ii) $v_t=\sfv_1,\theta_t=\sfv_1+\epsilon$ gives higher accrued value but consumes a disproportionately larger per-slot expected negative contribution to ROSC.
Thus, one way to understand Corollary \ref{cor:regretLB} is that for any $\cA$ that exactly satisfies the  ROSC and has $g_3>\sfc$ (a constant), must suffer a regret of $\Omega(T)$ with Input 2, since 
 with Input 2, having  $g_3>\sfc$, winning slots with $v_t=\sfv_2$ and $\theta_t=\sfv_2+\sfa+\sfb$ comes at a disproportionate (ROSC) cost of losing out on winning slots with $v_t=\sfv_1$ and $\theta_t=\sfv_1+\epsilon$ given that $\sfv_2/\sfv_1<2$.

Let the probability measure or expectation under Input $i, \ i=1,2$ be denoted as $\bbP^i$ or $\bbE^i$.
Consider any online algorithm $\cA$. Let the number of slots for which $\cA$ bids $b_t \ge \sfv_2+\sfb$ when $v_t = \sfv_2$ be $N_\cA$. Recall Definition \ref{defn:p3} for $g_1,g_2, g_3$. Let $\cE$ be the event that $N_\cA = o(T)$, i.e. $g_3=\frac{o(T)}{T}$.

Similar to \eqref{eq:consLBGen}, for $\cA$ to satisfy the ROSC with Input 1, we have
$$g_1 \frac{1}{4} \epsilon + g_2 \sfa \frac{1}{2} \left(r\right)  + g_3 (\sfa+\sfb) \frac{1}{2} \left(1-r\right) \le \frac{\epsilon}{4}.$$
Given that $g_3=\frac{o(T)}{T}$, we have 
$$g_1 \frac{1}{4} \epsilon + g_2 \sfa \frac{1}{2} \left(r\right)  +  \frac{o(T)}{T}(\sfa+\sfb) \frac{1}{2} \left(1-r\right) \le \frac{\epsilon}{4}.$$
Moreover, by definition $\sfa \ge \frac{2\epsilon}{3}$. Thus, we get 
\begin{equation}\label{eq:input1efrosc}
g_1 \frac{1}{4} \epsilon + g_2 \frac{2\epsilon}{3} \frac{1}{2} r  \le \frac{\epsilon}{4},
\end{equation}
since 
$\frac{o(T)}{T}(\sfa+\sfb)\frac{1}{2} \left(1-r\right)\ge 0$.
Moreover, the expected accrued valuation of $\cA$ with Input 1 is 
$$\left(\frac{\sfv_1 T}{4} + \frac{g_1\sfv_1 T}{4} + \frac{\sfv_2 g_2 T}{2} + \sfv_2 (1-r) o(T)\right).$$
Enforcing \eqref{eq:input1efrosc} and recalling that $r< \frac{1}{2}$ implies that 
the expected accrued valuation of $\cA$ with Input 1 is at most
$$\left(\frac{\sfv_1 T}{4} + \frac{c \sfv_2T}{4} + \sfv_2 (1-r) o(T)\right),$$
since $\sfv_2=0.9$ and $\sfv_1=0.5$ for some constant $c<1$.



Recall that with Input 1, the total accrued valuation for $\opt$ is at least
$\frac{\sfv_1 T}{4} + \frac{\sfv_2 T}{4}$.
Thus, when event $\cE$ happens ($N_\cA = o(T)$),
$$\cR_\cA^1(T) \ge \left(\frac{\sfv_2 (1-c) T}{4} \right) \bbP^1(\cE) > \Omega(T)\bbP^1(\cE).$$

However, if $\cE^c$ is true, then $g_3>\sfc_1$ (for some constant $\sfc_1$) for $\cA$, in which case Corollary \ref{cor:regretLB} implies that either the ROSC is violated by $\cA$ or the 
regret of $\cA$ is at least $\Omega(T)$. Thus, for $\cA$ that satisfies the ROSC, we have $$\cR^2_\cA(T) \ge \Omega(T) \bbP^2(\cE^c).$$

From the Bretagnolle-Huber inequality \cite{lattimore2020bandit}
 we have 
$$\bbP^1(\cE) + \bbP^2(\cE^c) \ge \frac{1}{2} \exp\left(-\textsf{KL}(\bbP^1|| \bbP^2)\right) \stackrel{(a)}\ge \frac{1}{2} \exp\left(- 2T c_2\delta^2\right),$$
where $(a)$ follows from simple computation given the choice of Input 1 and Input 2 and $c_2$ is a constant.

Therefore, we get 
$$\cR_\cA^1(T) + \cR_\cA^2(T) \ge \Omega(T) \exp\left(- 2c_2T \delta^2\right).$$
Choosing $\epsilon =1/\sqrt{T}$ and since $\delta=\Theta(\epsilon)$, we get that 
$\cR_\cA^1(T) + \cR_\cA^2(T) \ge \Omega(T).$
\end{proof}
\section{Proof of Theorem \ref{thm:lbuniv}}\label{app:lbequal}
\begin{proof}
Consider Input 1: $v_t=\frac{1}{2}$ $\forall \ t$, and allocation functions are $x_t^{\theta_t}$  \eqref{def:thresholdfunc}, where $\theta_t$ takes four possible values $A=\frac{1}{2}-u, B=\frac{1}{2}-w, C=\frac{1}{2} + m_1 - \delta, D=\frac{1}{2} + m_1$, with equal probability of $1/4$ where $0< w< u < \frac{1}{2}$, $m_1 = \frac{\frac{1}{2}-u + \frac{1}{2}-w}{2}$ (the conditional mean of $\theta_t$ given $\theta_t \in \{A,B\}$) and $\delta=\frac{1}{\sqrt{T}}$.

Similarly let Input 2: $v_t=\frac{1}{2}$ $\forall \ t$, and allocation functions are $x_t^{\theta_t}$ \eqref{def:thresholdfunc}, where $\theta_t$ takes the same four values as in Input 1, $A,B,C,D$, but with perturbed probabilities. 
In particular, $\bbP(\theta_t\in \{A, B\}) =\frac{1}{2}$, but the marginal probabilities of $\{A, B\}$ are chosen such that the conditional mean of $\theta_t$ given $\theta_t \in \{A,B\}$ is $m_1+\delta$, while 
$\theta_t\in\{C, D\}$ are chosen with probability $1/4$ each as with Input 1. 

See Fig. \ref{fig:inputrepauction} for a pictorial description where quantities with blue and red color correspond to same sign on their ROSC contribution.
\begin{figure}
\includegraphics[width=10cm,keepaspectratio,angle=0]{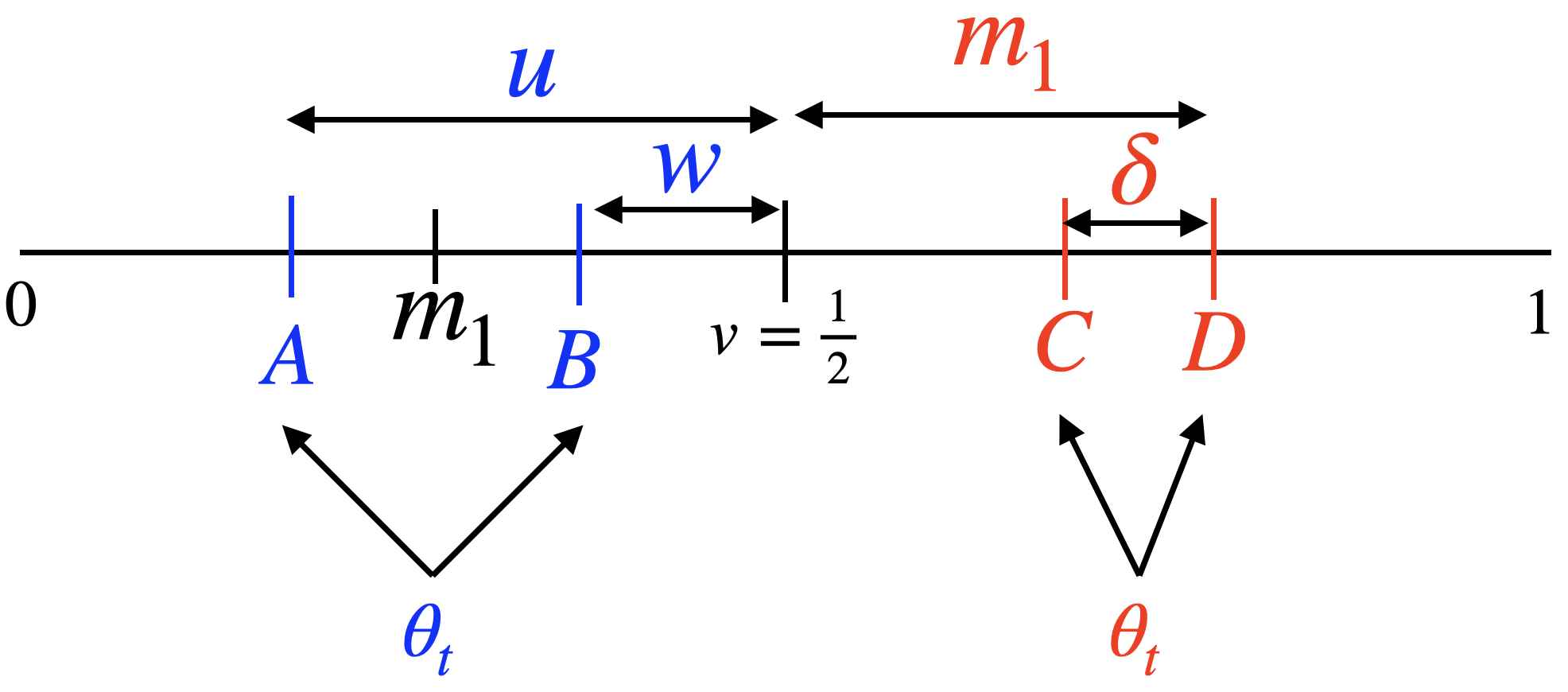}
\caption{Input used to prove Theorem \ref{thm:lbuniv}.}
\label{fig:inputrepauction}
\end{figure}

With both Input 1 and Input 2, $(v_t=1/2, \theta_t)$ is realized i.i.d. in each slot $t$ with distribution as described above. 

Consider the behaviour of $\opt$. If the input is Input 1, $\opt$ bids $b_t = D$ for all slots and wins all of them while satisfying the ROSC, since the expected positive (negative) per-slot contribution to ROSC when $\theta \in \{A,B\}$ ($\theta \in \{C,D\}$) is $m_1$ ($\ge -m_1$). 
Consequently, the total  accrued valuation by $\opt$ is $v\cdot T = T/2$, while satisfying the ROSC.

 In contrast, if the input is Input 2, then consider an algorithm $\cB$ (that need not be $\opt$) that always bids $b_t = C$. Clearly, $\cB$ satisfies the ROSC, thus so can $\opt$.

Given that the input is either Input 1 and Input 2, WLOG, we let any $\cA$ only bid either $b_t = C$ or 
$b_t=D$ for all $t$. 
\begin{lemma}\label{lem:ccvC}
For any slot $t$, if  $\cA$ bids $b_t=C$, then under  Input 2, $$\bbE\{\text{CV}_{\cA}(t)\}= -\left(\frac{1}{4}-\frac{3m_1}{4}-\frac{\delta}{4}\right).$$
With $m_1=\frac{1}{4}$, $\bbE\{\text{CV}_{\cA}(t)\}=-(\frac{1}{16}-\frac{\delta}{4})$.
\end{lemma}
\begin{proof}
Whenever  $\cA$ bids $b_t=C$ and let under Input 2. if $\theta_t =D$, then  $\text{CV}_{\cA}(t)=0$ since $\cA$ loses that slot and $p_t^{\theta_t}=0, v_tx_t^{\theta_t}(b_t)=0$.
Thus, the only cases to consider are when $\theta_t \in \{A,B,C\}$, and for which we get $\bbE\{\text{CV}_{\cA}(t)\} =$
$ - \frac{1}{4} \left(\frac{1}{2} - C\right) - \frac{1}{2}\left(\frac{1}{2}-(m_1+\delta)\right) = -\left(\frac{1}{4}-\frac{3m_1}{4}-\frac{\delta}{4}\right)$,
where the last equality is obtained by plugging in the value for $C$. 
Let $m_1=\frac{1}{4}$, then $\bbE\{\text{CV}_{\cA}(t)\}=-(\frac{1}{16}-\frac{\delta}{4})$ when $\cA$ chooses $b_t=C$ for any slot $t$.
\end{proof}

\begin{lemma}\label{lem:ccvD}
For any slot $t$, if $\cA$ bids $b_t=D$, and then under Input 2, then $\bbE\{\text{CV}_{\cA}(t)\} =\frac{\delta}{4}.$
\end{lemma}
\begin{proof}
With Input 2, if $\cA$ chooses $b_t=D$ in slot $t$, all possible values of $\theta_t \in \{A,B,C,D\}$ result in non-trivial CV. 
We compute the expected $\text{CV}$ as follows.  
\begin{align}\nn
\bbE\{\text{CV}_{\cA}(t)\} &= -\frac{1}{4} \left(\frac{1}{2} - C\right) - \frac{1}{4} \left(\frac{1}{2} - D\right) -   \frac{1}{2}\left(\frac{1}{2}-(m_1+\delta)\right) = \frac{\delta}{4}.
\end{align}
\end{proof}

\begin{lemma}\label{lem:ccvlb}
If the number of slots for which $\cA$ chooses $b_t=D$ is at least as much as $T-\sqrt{T}$, then if the input is Input 2, 
$\bbE\{\text{CCV}_{\cA}(T)\} \ge \frac{\sqrt{T}}{16} - \frac{T^{1/4}}{4}$ for $\delta=\frac{1}{\sqrt{T}}$.

\end{lemma}
\begin{proof}  $\bbE\{\text{CCV}_{\cA}(T)\} = \sum_{t=1}^{T} \bbE\{\text{CV}_{\cA}(t)\}$
$$\stackrel{(a)}\ge -\left(\frac{1}{16}-\frac{\delta}{4}\right)\sqrt{T} + \frac{\delta}{4} (T-\sqrt{T}) \stackrel{(b)} \ge  \frac{\sqrt{T}}{16} - \frac{T^{1/4}}{4},$$
where $(a)$ follows by combining Lemma \ref{lem:ccvC} and \ref{lem:ccvD} if the number of slots for which $b_t=D$ is at least as much as $T-\sqrt{T}$, while $(b)$ follows by plugging in $\delta=\frac{1}{\sqrt{T}}$.
\end{proof}

Let the probability measure or expectation under Input $i, \ i=1,2$ be denoted as $\bbP^i$ or $\bbE^i$.
Let 
$$\cE = \{\# \text{of slots $t$ for which} \ b_t=C \ \text{by} \ \cA > T-\sqrt{T}\}.$$
Recall that with Input $1$, $\theta_t=D$ with probability $1/4$. Therefore all slots, where $\cA$ chooses $b_t=C$ and for which $\theta_t=D$, are lost by $\cA$. Recall that with Input 1, the total accrued reward of $\opt$ is $T/2$. 
Thus, the regret of $\cA$ is at least 
$$\cR_{\cA}(T) \ge \frac{\sqrt{T}}{4} \bbP^1(\cE).$$

As pointed before, any online algorithm $\cA$ will only bid $b_t = C$ or 
$b_t=D$ for all $t$ given the definition of Input 1 and 2, i.e., either event $\cE$ happens or its complement $\cE^c$. 
Thus, from Lemma \ref{lem:ccvlb}, the constraint violation for $\cA$ is 
$$\bbE\{\text{CCV}_\cA(T)\} \ge \left(\frac{\sqrt{T}}{16} - \frac{T^{1/4}}{4}\right)\bbP^2(\cE^c).$$

From the Bretagnolle-Huber inequality \cite{lattimore2020bandit}
 we have 
$$\bbP^1(\cE) + \bbP^2(\cE^c) \ge \frac{1}{2} \exp\left(-\textsf{KL}(\bbP^1|| \bbP^2)\right) \stackrel{(a)}\ge \frac{1}{2} \exp\left(- 2T \delta^2\right),$$
where $(a)$ follows from the choice of Input 1 and Input 2.

Therefore, we get 
$$\cR_{\cA}(T) + \bbE\{\text{CCV}_{\cA}(T)\} \ge \frac{\sqrt{T}}{16} \exp\left(- 2T \delta^2\right) - o(T).$$
Since $\delta=1/\sqrt{T}$, we get the result.
\end{proof}
\section{Proof of Theorem \ref{thm:ubequalcase}}\label{app:thm:ubequalcase}
{\bf Case I $\mu_L + \mu_R\le 0$}
\begin{prop}\label{prop:crossingtime}
$\bbE\{\tau_{\min}\}=O(\sqrt{T})$ if
$\Delta = \mu_L+ \underline {\mu}_R^{\theta^\star} = \mu_L+ \bbE\{\b1_{v< \theta}\cdot (v-\theta) | \theta<\theta^\star\}>0$ does not depend on $T$.
\end{prop}
\begin{proof}
From Definition \ref{defn:mindrift}, $\text{Margin}(t)$ process  for $\cA_c$ has  drift $\Delta >0$ until $b_t<\theta^\star$ (and consequently the $b_t$ process has drift $\Delta/\sqrt{T}$). Moreover, since $\theta^\star\le 1$, following a standard 
result from random walks with positive drift \cite{feller1991introduction}, we have $\bbE\{\tau_{\min}\}=O(\frac{\sqrt{T}}{\Delta}) = O(\sqrt{T})$.
\end{proof}

Proposition \ref{prop:crossingtime} implies the following Lemma.
 \begin{lemma}\label{lem:regretcrossingtime}
$\cR_{\cA_c}([1:\tau_{\min}])=O(\sqrt{T})$ when $\Delta$ does not depend on $T$.
\end{lemma}

Next, we characterize  $\cR_{\cA_c}(\cT^{-}_{\theta^\star})$ 
by establishing a concentration bound for $b(t)$ as follows.
\begin{lemma}\label{lem:coupling}
For $\cA_c$, at any $t\ge \tau_{\min}$,  $\bbP(b_t < (1-\delta)\theta^\star) \le \exp\left(-  \delta \Delta \theta^\star \sqrt{T} c\right)$, where $c$ is a constant that depends on the distribution of allocation function \eqref{def:thresholdfunc}.
\end{lemma}

\begin{proof}
Consider the following  process
\begin{equation}\label{eq:alg2}
N(t) = \max\{N(t-1) + A(t) - B(t),0\}
\end{equation} and 
\vspace{-0.1in}
\begin{equation}\label{eq:alg1}
B(t) = \begin{cases} 
\mu - \nu^{-} & \text{if} \  N(t) \le U/2,\\
 \mu + \nu^{+} &\text{if} \ N(t) > U/2,
\end{cases}
\end{equation} 
where $A(t)\ge 0$ is a ``well-behaved" arrival process\footnote{Asymptotic semi-invariant log moment generating function exists for $(-\inf, r)$ for some $r>0$.} with $\bbE\{A(t)\}=\mu$ and $U >0$ is some threshold, and $0< \nu^+<\nu^-  < \mu$.  
Essentially, $N(t)$ can be thought of as a queue length process where arrivals $A(t)$ are exogenous, and departures $B(t)$ are regulated by a controller with an objective of keeping  $N(t)$ close to a threshold of $U/2$ as much as possible. To achieve this, the control $B(t)$ is chosen to be little more (less) than the expected arrival rate $\mu$ whenever $N(t)$ is above (below) the threshold $U/2$. The upward drift whenever $N(t) \le U/2$ in \eqref{eq:alg1} is $\nu^{-}$ and downward drift is $\nu^+$ when $N(t) \le U/2$.

In \cite[Lemma 2]{Koksal} show that 
\begin{equation}\label{eq:resultkoksal}
\bbP(N(t) =0) \le \exp\left(- \nu^{-} U c\right),
\end{equation} for any  $U$, and where $c$ depends on the 
distribution of $A(t)$.

To derive our result we make the following connection between $\text{Margin}(t)$ process \eqref{defn:marginprocess} and \eqref{eq:alg2}.
The $\text{Margin}(t)$ process \eqref{defn:marginprocess} (that drives the bid process \eqref{eq:bidprocess}) is similar to \eqref{eq:alg2}, where the possible increments/decrements at slot $t$ are 
$(v_t-\theta_t)$ (which can be positive/negative) only if $\text{Margin}(t-1)\ge \sqrt{T}(v_t-\theta_t)$. Thus, $\sqrt{T} \theta^\star$ in \eqref{defn:marginprocess} is analogous to the threshold $U/2$ in \eqref{eq:alg2}, and the increments/decrements $(v_t-\theta_t)$  are analogous 
to $A(t)-B(t)$ in \eqref{eq:alg2}, and the enforced drift $\nu^{-}$ in \eqref{eq:alg1} below $U/2$  is now $\mu_L+ \underline {\mu}_R^{\theta^\star}=\Delta > 0$ when the $\text{Margin}$ process is below $\sqrt{T} \theta^\star$ for \eqref{eq:bidprocess}) and $\nu^+$ is $\mu_L+\bar {\mu}_R^{\theta^\star}$ when the $\text{Margin}$ process is above $\sqrt{T} \theta^\star$.
From Assumption \ref{assump:updown}, $|\mu_L+\bar {\mu}_R^{\theta^\star}| < |\Delta|$. Thus,  following an identical proof to \cite[Lemma 2]{Koksal} used to derive \eqref{eq:resultkoksal}, 
we get
$$\bbP(b(t) < (1-\delta)\theta^\star) \le \exp\left(-  \delta \Delta \theta^\star \sqrt{T} c\right).$$
\end{proof} 

Using Lemma \ref{lem:coupling}, next, we upper bound  $ \cR_{\cA_c}(\cT^{-}_{\theta^\star})$.
Choose $\delta=\sqrt{\frac{\log (T)}{T}}$. Lemma \ref{lem:coupling} implies that $\cA_c$ wins any slot belonging to $\cT^{-}$ for which $\theta < (1-\delta)\theta^\star$ with probability at least $1-\frac{c_1}{\sqrt{T}}$ when $\Delta$ is independent of $T$.
Thus, using the linearity of expectation, the expected number of slots that $\cA_c$ wins among $\cT^{-}_{\theta^\star}$ for which $\theta < (1-\delta)\theta^\star$ is 
$\left(1-\frac{c_1}{\sqrt{T}}\right)|\cT^{-}_{\theta^\star}|$. Moreover, for the interval, $[(1-\delta)\theta^\star, \theta^\star)$, let $T$ be large enough such that 
the total probability mass of $\theta \in  [(1-\delta)\theta^\star, \theta^\star)$ be at most $\delta=\sqrt{\frac{\log (T)}{T}}$, since $\theta^\star \le 1$. Thus,  
the expected number of slots that $\cA_c$ can lose among $\cT^{-}_{\theta^\star}$ for which $[(1-\delta)\theta^\star, \theta^\star)$ is 
$T \cdot \delta = T \sqrt{\frac{\log (T)}{T}} = \sqrt{T \log T}$.
Thus, 
\begin{equation}\label{eq:dummy11}
\cR_{\cA_c}(\cT^{-}_{\theta^\star}) = O(\sqrt{T \log T}).
\end{equation}

 Next, we identify the rate at which the $b_t$ process of $\cA_c$ crosses the threshold $\theta^\star$ \eqref{defn:thetastar} as a function of $\pi^\star$ \eqref{defn:thetastar}.
  \begin{lemma}\label{lem:hittingthetastar}
For $t\ge \tau_{\min} $, 
for $b_t$ \eqref{eq:bidprocess}, $\bbP(b_t\ge \theta^\star) \ge \rho^\star$, where $\rho^\star \bbP(\theta=\theta^\star) = \pi^\star$.
\end{lemma}
\begin{proof}
Let the claim be false, i.e, $\bbP(b_t\ge \theta^\star)< \rho^\star$. Then the probability of winning any slot by $\cA_c$ when $\theta=\theta^\star$ is   $\bbP(b_t\ge \theta^\star)\bbP(\theta=\theta^\star)< \rho^\star \bbP(\theta=\theta^\star) < \pi^\star$. 
But from the definition \eqref{defn:thetastar} of $\theta^\star$ and $\pi^\star$ when $\bbP(b_t\ge \theta^\star)\bbP(\theta=\theta^\star)< \pi^\star$, the $\text{Margin}(t)$ process \eqref{defn:marginprocess} and effectively the $b_t$ process \eqref{eq:bidprocess} of $\cA_c$ has positive drift\footnote{The expected  increase minus the expected decrease in any slot.}
 even when $b_t=\theta^\star$. Thus, similar to Lemma \ref{lem:coupling}, we will get that with high probability $b_t > \theta^\star$ for $t\ge \tau_{\min}$.  However, given that the algorithm $\cA_c$ satisfies the ROSC on a sample path basis,  this contradicts the definition of $\theta^\star$, since $\theta^\star$ is the highest bid that $\opt$ can make with non-zero probability while satisfying the ROSC. 
 Thus, we get a contradiction to the claim that $\bbP(b_t\ge \theta^\star)< \rho^\star$.
\end{proof}

Next, using Lemma \ref{lem:hittingthetastar}, we show that $\cR_{\cA_c}( \cT_{\theta^\star})=0$.
\begin{lemma}\label{lem:RAstar}
  $\cR_{\cA_c}( \cT_{\theta^\star})=0.$
\end{lemma}
\begin{proof}\nonumber
\begin{align} \cR_{\cA_c}( \cT_{\theta^\star})  & = \bbE\{\#\text{slots won by} \ \opt \ \text{in} \ \cT_{\theta^\star}\}    -  \bbE\{\#\text{slots won by} \  \cA_c \ \text{in} \ \cT_{\theta^\star} \}, \\
&  \stackrel{(a)}\le  0,
\end{align}
where $(a)$ follows from Lemma \ref{lem:hittingthetastar} that shows that both $\cA_c$ and $\opt$ win slots with $\theta=\theta^\star$ with probability $\pi^\star$.
\end{proof}

\begin{proof}[Proof of Theorem \ref{thm:ubequalcase}]
When  $\Delta$ does not depend on $T$, combining  Lemma \ref{lem:regretcrossingtime},  Lemma \ref{lem:RAstar}, and \eqref{eq:dummy11} and recalling that  $\cR_{\cA_c}(\cT^+_{\theta^\star})\le 0$, we get 
$$\cR_{\cA}(T) = \cR_{\cA}([1:\tau_{\min}]) + \cR_{\cA_c}(\cT^+_{\theta^\star}) + \cR_{\cA_c}( \cT_{\theta^\star})+  \cR_{\cA_c}(\cT^{-}_{\theta^\star}) \le O(\sqrt{T \log T}).$$ 

{\bf Case II  $\mu_L + \mu_R>0$} 
In this case, $\opt$ wins all slots since satisfying the ROSC  is trivial by bidding $b_t^\opt=1$ for all $t$.
For $\cA_c$, let $\tau'_{\min}= \min\{t: b_t \ge 1\}.$ Since $\mu_L + \mu_R>0$, the $\text{Margin}(t)$ process \eqref{defn:marginprocess} for $\cA_c$ has a positive drift for all slots. Thus, similar to Lemma \ref{lem:regretcrossingtime}, $\bbE\{\tau'_{\min}\} = O(\sqrt{T})$, and similar to 
Lemma \ref{lem:coupling}, with high probability $b_t \ge 1$ for $t\ge \tau'_{\min}$. Thus, similar to Case I, we get  $\cR_{\cA_c}(T) = O(\sqrt{T \log T}).$
\end{proof}

\section{Proof of Lemma \ref{lem:order}}
\begin{proof}
Without loss of generality, let all $\frac{ \sfv_k}{r_k}$ be distinct.
  Let the statement be false, and in particular in the optimal solution $\bq$ there exists a $k'$ (in order) for which $q_{k'} < 1$  but $q_{k'+1} > 0$. We will contradict to the optimality of $\bq$ by creating a new solution with larger objective function while satisfying the constraint. Consider a new solution ${\hat \bq}$ where we keep all 
  $q_k$'s the same other than $q_{k'}$ and $q_{k'+1}$ which are changed to ${\hat q}_{k'} = q_{k'}+q_{k'+1} \frac{r_{k'+1}}{r_{k'}}$ and ${\hat q}_{k'+1}=0$. Clearly, ${\hat \bq}$ satisfies the constraint. However, the change in objective function value (going from $\bq$ to $\hat \bq$) is 
  $$q_{k'+1}r_{k'+1} \left( \frac{\sfv_{k'}}{r_{k'}} - \frac{\sfv_{k'+1}}{r_{k'+1}}\right) >0,$$ since $\frac{\sfv_{k'}}{r_{k'}} > \frac{\sfv_{k'+1}}{r_{k'+1}}$ by definition.
  Thus, $\bq$ cannot be optimal.
\end{proof}

\section{Proof of Theorem \ref{proof:learn1}}
\begin{proof}
Since all the parameters of $\cP_{\text{light}}$ are expected values of i.i.d. random variables that are estimated using sample mean estimator $\frac{1}{n}\sum_{i=1}^n X_i$, from Chernoff bound, we have 
$ \bbP\left( \left | \frac{1}{n}\sum_{i=1}^n X_i - \bbE\{X_1\} \right | > \epsilon\right) \le \exp^{ -(2n \epsilon^2)}$.
Thus, we get that dedicating $T/2$ slots for learning, the probability that any of parameters of $\cP_{\text{light}}$ deviate from their means by more than $O(\log(1/\delta)/\sqrt{T})$ is at most $K \delta$ (union bound).

Once the learning is complete at the end of the $T/2^{th}$ slot, \textsf{Learn-Alg} employs the optimal solution (Lemma \ref{lem:order}) assuming all the parameters 
of $\cP_{\text{light}}$ are known exactly. However, since the parameters of $\cP_{\text{light}}$ are learnt, there is always a residual error, and hence we appeal next to the sensitivity analysis \cite{boyd2004convex} of LPs, which deals with the effect of perturbation of parameters/constraints on the optimal LP solution.
From [Section 5.6.3 \cite{boyd2004convex}], for LPs (where strong duality holds, which is true in our case since \eqref{eq:optprob} is feasible and bounded), choosing $\delta = 1/T$, we get that  the solution obtained by \textsf{Learn-Alg} satisfies
  $$\frac{ \bbE\left\{\ \sum_{t=T/2+1}^T v_t x_t(b^\cA_t)\right\} }{ \bbE\left\{\ \sum_{t=T/2+1}^T v_t x_t(b^\opt_t)\right\} }\ge (1-O(K\log(1/\delta)/\sqrt{T})) ,$$ 
  since when random variables are well estimated within an error of $O(\log(1/\delta)/\sqrt{T})$, the value accrued by algorithm \textsf{Learn-Alg} is within $(1-O(K\log(1/\delta)/\sqrt{T}))$ fraction of that of $\opt$, while otherwise, 
 the value accrued by algorithm \textsf{Learn-Alg} is at least $0$.
  
  Since the input is i.i.d., $\bbE\left\{\ \sum_{t=1}^T v_t x_t(b^\opt_t)\right\} = 2 \bbE\left\{\ \sum_{t=T/2+1}^T v_t x_t(b^\opt_t)\right\}.$
  Thus, we get that $\text{c.r.}_{\text{\textsf{Learn-Alg}}} \ge 1/2 -(1-O(K\log(1/\delta)/\sqrt{T}))$.
  Recall that until slot $T/2$, $b_t=v_t$, thus \textsf{Learn-Alg} satisfies the ROSC until slot $T/2$ following Section \ref{sec:warmup}.
  Moreover, for $t>T/2$, the solution output by \textsf{Learn-Alg}  satisfies the ROSC with probability $1-(1-O(K\log(1/\delta)/\sqrt{T}))$. 
\end{proof}

 	\end{document}